\documentclass[conference,hidelinks]{IEEEtran}
\IEEEoverridecommandlockouts
\usepackage{cite}
\usepackage{amsmath,amssymb,amsfonts}
\usepackage{algorithmic}
\usepackage{graphicx}
\usepackage{textcomp}
\usepackage{xcolor}
\usepackage{orcidlink}
\usepackage{glossaries}
\usepackage[position=b]{subcaption}
\usepackage[a4paper, total={184mm,239mm}]{geometry}
\usepackage{enumitem}
\usepackage{hyperref}
\usepackage{cleveref}
\usepackage[scaled=0.95]{inconsolata}
\usepackage[outputdir=build,frozencache]{minted}

\def\BibTeX{{\rm B\kern-.05em{\sc i\kern-.025em b}\kern-.08em
    T\kern-.1667em\lower.7ex\hbox{E}\kern-.125emX}}

\makeatletter
\newcommand{\insertnewlines}[1]{%
  \noindent\mbox{}%
  \@tempcnta=#1\relax
  \loop\ifnum\@tempcnta>0
    \\
    \advance\@tempcnta\m@ne
  \repeat
}
\makeatother

\newacronym[plural=WANs, firstplural={Wide Area Networks (WANs)}]{wan}{WAN}{Wide Area Network}
\newacronym[plural=WSNs, firstplural={Wireless Sensor Networks (WSNs)}]{wsn}{WSN}{Wireless Sensor Network}
\newacronym{simd}{SIMD}{Single Instruction Multiple Data}
\newacronym{os}{OS}{Operating System}
\newacronym{ble}{BLE}{Bluetooth Low-Energy}
\newacronym{wifi}{Wi-FI}{Wireless Fidelity}
\newacronym[plural=DVS, firstplural={Dynamic Vision Sensors (DVS)}]{dvs}{DVS}{Dynamic Vision Sensor}
\newacronym{ptz}{PTZ}{Pan-Tilt Unit}

\newacronym[plural=FLLs,firstplural=Frequency Locked Loops (FLLs)]{fll}{FLL}{Frequency Locked Loop}
\newacronym{dram}{DRAM}{Dynamic Random Access Memory}
\newacronym{fpu}{FPU}{Floating Point Unit}
\newacronym{fpss}{FPSS}{Floating Point Subsystem}
\newacronym{frep}{FREP}{Floating Point Repetition}
\newacronym{dma}{DMA}{Direct Memory Access}
\newacronym{ssr}{SSR}{Stream Semantic Register}
\newacronym{issr}{ISSR}{Indirection Stream Semantic Register}
\newacronym[plural=LUTs, firstplural={Lookup Tables (LUTs)}]{lut}{LUT}{Lookup Table}
\newacronym[plural=FPGAs, firstplural={Field Programmable Gate Arrays (FPGAs)}]{fpga}{FPGA}{Field Programmable Gate Array}
\newacronym{dsp}{DSP}{Digital Signal Processing}
\newacronym{mcu}{MCU}{Microcontroller Unit}
\newacronym{spi}{SPI}{Serial Peripheral Interface}
\newacronym{cpi}{CPI}{Camera Parallel Interface}
\newacronym{rf}{RF}{register file}
\newacronym{rs}{\texttt{rs}}{source register}
\newacronym{rd}{\texttt{rd}}{destination register}
\newacronym{fifo}{FIFO}{First In First Out}
\newacronym{uart}{UART}{Universal Asynchronous Receiver-Transmitter}
\newacronym{raw}{RAW}{Read After Write}
\newacronym[plural=ISAs, firstplural={Instruction Set Architectures (ISAs)}]{isa}{ISA}{Instruction Set Architecture}
\newacronym{xbar}{XBAR}{crossbar}
\newacronym[firstplural=Scratch-Pad Memories (SPMs)]{spm}{SPM}{Scratch-Pad Memory}
\newacronym{ppa}{PPA}{Power Performance Area}
\newacronym{ipi}{IPI}{Inter-Processor Interrupt}
\newacronym[firstplural=Software-Generated Interrupts (SGIs)]{sgi}{SGI}{Software-Generated Interrupt}
\newacronym{pe}{PE}{Processing Element}
\newacronym{tcdm}{TCDM}{Tightly-Coupled Data Memory}
\newacronym{lsu}{LSU}{Load-Store Unit}
\newacronym{icache}{I\$}{Instruction Cache}
\newacronym{dcache}{D\$}{Data Cache}
\newacronym{wfi}{WFI}{Wait For Interrupt}
\newacronym{gpc}{GPC}{GPU Processing Cluster}
\newacronym{cpu}{CPU}{Central Processing Unit}
\newacronym{gpu}{GPU}{Graphics Processing Unit}
\newacronym{llc}{LLC}{Last-Level Cache}
\newacronym{sm}{SM}{Streaming Multiprocessor}
\newacronym[firstplural=Networks on Chip (NoCs)]{noc}{NoC}{Network on Chip}

\newacronym{dfg}{DFG}{Data Flow Graph}
\newacronym{lcg}{LCG}{Linear Congruential Generator}
\newacronym{prn}{PRN}{Pseudo-Random Number}

\newacronym{ste}{STE}{Straight-Through-Estimator}

\newacronym[plural=PTUs, firstplural={Pan-Tilt Units}]{ptu}{PTU}{Pan-Tilt Unit}
\newacronym{mdf}{MDF}{Medium-density fibreboard}
\newacronym{cvat}{CVAT}{Computer Vision Annotation Tool}
\newacronym{coco}{COCO}{Common Objects in Context}
\newacronym{soa}{SoA}{State of the Art}
\newacronym{sf}{SF}{Sensor Fusion}

\newacronym{dl}{DL}{Deep Learning}
\newacronym{bn}{BN}{Batch Normalization}
\newacronym{FGSM}{FBK}{Fast Gradient Sign Method}
\newacronym{lr}{LR}{Learning Rate}
\newacronym{sgd}{SGD}{Stochastic Gradient Descent}
\newacronym{gd}{GD}{Gradient Descent}
\newacronym{llm}{LLM}{Large Language Model}

\newacronym{sta}{STA}{Static Timing Analysis}

\newacronym[plural=GPIOs, firstplural={General Purpose Inupt Outputs (GPIOs)}]{gpio}{GPIO}{General Purpose Input Output}
\newacronym[plural=LDOs, firstplural={Low Dropout Regulators (LDOs)}]{ldo}{LDO}{Low Dropout Regulator}

\newacronym{inq}{INQ}{Incremental Network Quantization}

\newacronym{CV}{CV}{Computer Vision}
\newacronym{EoT}{EoT}{Expectation over Transformation}
\newacronym{RPN}{RPN}{Region Proposal Network}
\newacronym{TV}{TV}{Total Variation}
\newacronym{NPS}{NPS}{Non-Printability Score}
\newacronym{STN}{STN}{Spatial Transformer Network}
\newacronym{MTCNN}{MTCNN}{Multi-Task Convolutional Neural Network}
\newacronym{YOLO}{YOLO}{You Only Look Once}
\newacronym{SSD}{SSD}{Single Shot Detector}
\newacronym{SOTA}{SOTA}{State of the Art}
\newacronym{NMS}{NMS}{Non-Maximum Suppression}
\newacronym{ic}{IC}{Integrated Circuit}
\newacronym{tcxo}{TCXO}{Temperature Controlled Crystal Oscillator}
\newacronym{jtag}{JTAG}{Joint Test Action Group industry standard}
\newacronym{swd}{SWD}{Serial Wire Debug}
\newacronym{sdio}{SDIO}{Serial Data Input Output}

\newacronym[plural=PCBs, firstplural={Printed Circuit Boards (PCB)}]{pcb}{PCB}{Printed Circuit Board}
\newacronym[plural=ASICs, firstplural={Application Specific Integrated Circuits}]{asic}{ASIC}{Application Specific Integrated Circuit}

\newacronym[plural=BNNs, firstplural={Binary Neural Networks (BNNs)}]{bnn}{BNN}{Binary Neural Network}
\newacronym[plural=NNs, firstplural={Neural Networks}]{nn}{NN}{Neural Network (NNs)}
\newacronym[plural=SCMs, firstplural={Standard Cell Memories (SCMs)}]{scm}{SCM}{Standard Cell Memory}
\newacronym{ann}{ANN}{Artificial Neural Networks}
\newacronym{ml}{ML}{Machine Learning}
\newacronym{ai}{AI}{Artificial Intelligence}
\newacronym{iot}{IoT}{Internet of Things}
\newacronym{fft}{FFT}{Fast Fourier Transform}
\newacronym[plural=OCUs, firstplural={Output Channel Compute Units (OCUs)}]{ocu}{OCU}{Output Channel Compute Unit}
\newacronym{alu}{ALU}{Arithmetic Logic Unit}
\newacronym{mac}{MAC}{Multiply-Accumulate}
\newacronym[firstplural={systems-on-chip (SoCs)}]{soc}{SoC}{system-on-chip}
\newacronym[firstplural={multi-processor systems-on-chip (MPSoCs)}]{mpsoc}{MPSoC}{multi-processor system-on-chip}

\newacronym{PGD}{PGD}{Projected Gradient Descend}
\newacronym{CW}{CW}{Carlini-Wagner}
\newacronym{OD}{OD}{Object Detection}

\newacronym{rrf}{RRF}{RADAR Repetition Frequency}
\newacronym{nlp}{NLP}{Natural Language Processing}
\newacronym{qam}{QAM}{Quadrature Amplitude Modulation}
\newacronym{rri}{RRI}{RADAR Repetition Interval}
\newacronym{radar}{RADAR}{Radio Detection and Ranging}
\newacronym{loocv}{LOOCV}{Leave-one-out cross validation}

\newacronym{bsp}{BSP}{Board Support Package}
\newacronym{ttn}{TTN}{The Things Network}
\newacronym{wip}{WIP}{Work in Progress}
\newacronym{json}{JSON}{JavaScript Object Notation}
\newacronym{qat}{QAT}{Quantization-Aware Training}

\newacronym{cls}{CLS}{Classification Error}
\newacronym{loc}{LOC}{Localization Error}
\newacronym{bkgd}{BKGD}{Background Error}
\newacronym{roc}{ROC}{Receiver Operating Characteristic}
\newacronym{frr}{FRR}{False Rejection Rate}
\newacronym{eer}{EER}{Equal Error Rate}
\newacronym{snr}{SNR}{Signal-to-Noise Ratio}
\newacronym{flop}{FLOP}{Floating-Point Operation}
\newacronym{fp}{FP}{floating-point}
\newacronym{fps}{FPS}{Frames Per Second}
\newacronym{oi}{OI}{Operational Intensity}
\newacronym[first={IPC (Instructions per Cycle)}]{ipc}{IPC}{Instructions per Cycle}

\newacronym{gsc}{GSC}{Google Speech Commands}
\newacronym{mswc}{MSWC}{Multilingual Spoken Words Corpus}
\newacronym{demand}{DEMAND}{Diverse Environments Multichannel Acoustic Noise Database}

\newacronym[plural=SNNs, firstplural={Spiking Neural Networks (SNNs)}]{snn}{SNN}{Spiking Neural Network}
\newacronym[plural=DNNs, firstplural={Deep Neural Networks (DNNs)}]{dnn}{DNN}{Deep Neural Network}
\newacronym[plural=TCNs,firstplural=Temporal Convolutional Networks]{tcn}{TCN}{Temporal Convolutional Network}
\newacronym[plural=CNNs,firstplural=Convolutional Neural Networks (CNNs)]{cnn}{CNN}{Convolutional Neural Network}
\newacronym[plural=TNNs,firstplural=Ternarized Neural Networks]{tnn}{TNN}{Ternarized Neural Network}
\newacronym{ds-cnn}{DS-CNN}{Depthwise Separable Convolutional Neural Network}
\newacronym{rnn}{RNN}{Recurrent Neural Network}
\newacronym{gcn}{GCN}{Graph Convolutional Network}
\newacronym{mhsa}{MHSA}{Multi-Head Self Attention}
\newacronym{crnn}{CRNN}{Convolutional Recurrent Neural Network}
\newacronym{clca}{CLCA}{Convolutional Linear Cross-Attention}

\newacronym{bf}{BF}{Beamforming}
\newacronym{anc}{ANC}{Active Noise Cancellation}
\newacronym{agc}{AGC}{Automatic Gain Control}
\newacronym{se}{SE}{Speech Enhancement}
\newacronym{mct}{MCT}{Multi-Condition Training}
\newacronym{mcta}{MCTA}{Multi-Condition Training \& Adaptation}
\newacronym{pcen}{PCEN}{Per-Channel Energy Normalization}
\newacronym{mfcc}{MFCC}{Mel-Frequency Cepstral Coefficient}
\newacronym{asr}{ASR}{Automated Speech Recognition}
\newacronym{kws}{KWS}{Keyword Spotting}
\newacronym{odl}{ODL}{On-Device Learning}

\newacronym{nl-kws}{NL-KWS}{Noiseless Keyword Spotting}
\newacronym{na-kws}{NA-KWS}{Noise-Aware Keyword Spotting}
\newacronym{odda}{ODDA}{On-Device Domain Adaptation}
\newacronym{hpm}{HPM}{High-Performance Mode}
\newacronym{lpm}{LPM}{Low-Power Mode}

\newcommand{\OldResultGeomeanIPCIncrease}{1.6}

\newcommand{\OldResultGeomeanEnergySaving}{1.3}

\newcommand{\ResultPeakSpeedup}{1.49}
\newcommand{\ResultPeakSpeedupOverSnitch}{1.96}
\newcommand{\ResultGeomeanSpeedup}{1.19}
\newcommand{\ResultPeakEnergySaving}{1.47}
\newcommand{\ResultPeakEnergySavingOverSnitch}{1.75}
\newcommand{\ResultGeomeanEnergySaving}{1.21}
\newcommand{\ResultPeakIPC}{1.81}

\crefname{step}{step}{steps}
\Crefname{step}{Step}{Steps}

\Crefname{figure}{Fig.}{Figs.}

\setminted[]{
    escapeinside=||,
    style=arduino,
    breaklines,
    fontsize=\footnotesize,
    baselinestretch=0.82
}

\setlist{noitemsep, topsep=0pt}

\glsunset{ipc}
\glsunset{gpu}
\glsunset{ml}
\glsunset{pe}
\glsunset{rs}
\glsunset{rd}
\glsunset{rf}

\begin{document}

\title{Late Breaking Results: Boosting Efficient Dual-Issue Execution on Lightweight RISC-V Cores}

\ifdefined\blindreview
\else
\author{
    \IEEEauthorblockN{Luca Colagrande}
    \IEEEauthorblockA{
        \textit{Integrated Systems Laboratory (IIS)} \\
        \textit{ETH Zurich}\\
        Zurich, Switzerland \\
        colluca@iis.ee.ethz.ch\orcidlink{0000-0002-7986-1975}
    }
    \and
    \IEEEauthorblockN{Luca Benini}
    \IEEEauthorblockA{
        \textit{Integrated Systems Laboratory (IIS)} \\
        \textit{ETH Zurich}\\
        Zurich, Switzerland \\
        lbenini@iis.ee.ethz.ch\orcidlink{0000-0001-8068-3806}
    }
}
\fi

\maketitle

\begin{abstract}
    Large-scale \gls{ml} accelerators rely on large numbers of \glspl{pe}, imposing strict bounds on the area and energy budget of each \gls{pe}.
    Prior work demonstrates that limited dual-issue capabilities can be efficiently integrated into a lightweight in-order open-source RISC-V core (Snitch), with a geomean \gls{ipc} boost of \OldResultGeomeanIPCIncrease$\times$ and a geomean energy efficiency gain of \OldResultGeomeanEnergySaving$\times$, obtained by concurrently executing integer and \gls{fp} instructions. Unfortunately, this required a complex and error-prone low level programming model (COPIFT).
    We introduce COPIFTv2 which augments Snitch with lightweight queues enabling direct, fine-grained communication and synchronization between integer and \gls{fp} threads.
    By eliminating the tiling and software pipelining steps of COPIFT, we can remove much of its complexity and software overheads.
    As a result, COPIFTv2 achieves up to a \ResultPeakSpeedup$\times$ speedup and a \ResultPeakEnergySaving$\times$ energy-efficiency gain over COPIFT, and a peak \gls{ipc} of \ResultPeakIPC.
    Overall, COPIFTv2 significantly enhances the efficiency and programmability of dual-issue execution on lightweight cores.
    Our implementation is fully open source and performance experiments are reproducible using free software.
\ifdefined\blindreview
\footnote{\url{https://hidden-for-double-blind-review.com}}
\else
\footnote{\url{https://github.com/pulp-platform/snitch_cluster/tree/7c2bdd9}}
\fi
\end{abstract}


\section{Introduction}

Energy and area efficiency are key constraints in the design of modern \gls{ml} accelerators integrating large numbers of \glspl{pe}. 
In this context, single-issue in-order cores are common, but recent research shows that limited multiple-issue capabilities can be integrated with modest area and power overheads \cite{colagrande2025, burgess2020, kra2024, patsidis2018, wygrzvwalski2024}.

Nvidia's Turing architecture \cite{burgess2020}, for example, implements concurrent FP32 and INT32 execution within the strict design constraints of \gls{gpu} stream processors. Its designers report an average 36\% throughput increase across gaming workloads; however, due to the architecture's proprietary nature, design details and associated area and power costs remain undisclosed.

Similarly, Zaruba et al. \cite{zaruba2021} present Snitch, a single-issue in-order RISC-V processor capable of concurrently executing INT32 and FP64 operations.
In their design, an integer control core fetches and issues instructions in order, while offloading \gls{fp} operations to a dedicated coprocessor, or \gls{fpss}.
When these operations are part of a hardware loop, the \gls{fpss} can buffer them after they are offloaded for the first time by the control core.
The \gls{fpss} can then independently issue these instructions for successive loop iterations, allowing the integer core to fetch and execute other instructions in parallel.
This design requires the integer and \gls{fp} threads to be completely independent.
While this greatly simplifies the hardware---no dependency handling or communication is required between the two pipelines---it limits the applicability of the approach to a narrow set of workloads, and the amount of parallelism that can be extracted from them: authors report a maximum \gls{ipc} of 1.16, well below the theoretical limit of 2.

\begin{figure}[t]
    \centering
    \includegraphics[width=0.97\columnwidth]{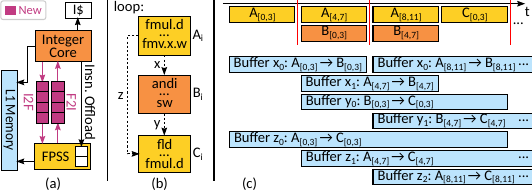}
    \begingroup
        \phantomsubcaption
        \label{fig:snitch}
        \phantomsubcaption
        \label{fig:kernel}
        \phantomsubcaption
        \label{fig:copift}
    \endgroup
    \caption{(a) Snitch architecture; (b) Integer and FP phases of the \texttt{exp} kernel; (c) COPIFT schedule and lifetime of buffers.}
\end{figure}

To overcome this limitation, Colagrande et al. \cite{colagrande2025} developed the COPIFT methodology.
Consider the computation in \Cref{fig:kernel}, mixing integer and \gls{fp} instructions.
COPIFT applies a sequence of code transformations to batch the computation and overlap independent integer and \gls{fp} batches.
Register-level inter-thread communication is spilled to memory, and dependencies are preserved by explicitly synchronizing the two threads in software at batch granularity.
The authors demonstrate the approach on mixed integer and \gls{fp} workloads, reporting up to 1.75 \gls{ipc}.

Despite its benefits, the methodology is complex and hinders programmability, introducing tradeoffs such as batch-size selection that require manual, workload-specific tuning. 
In addition, spilling all communication to memory introduces load/store instructions that consume additional issue cycles and energy.

In this work, we propose COPIFTv2, an enhanced version of the COPIFT methodology leveraging lightweight queues for direct, fine-grained communication between the integer and \gls{fp} threads.
We eliminate memory-based communication and batch-level synchronization, significantly improving both the energy-efficiency and programmability of COPIFT.

\section{Implementation}


As shown in \Cref{fig:snitch}, we augment Snitch with two lightweight queues, one from the integer core to the \gls{fpss} (I2F), and one in the opposite direction (F2I).
The queues provide: 1) a path for inter-thread communication, and 2) a mechanism for fine-grained inter-thread synchronization.
The latter is achieved through the blocking, FIFO semantics of the queues: pop (or read) operations must follow corresponding push (or write) operations, and stall if the queue is respectively empty or full.

To enable communication via the queues, we introduce a custom CSR (\texttt{EnCopiftQueues}), which alters source (\gls{rs}) and destination (\gls{rd}) register semantics in the following way:
\begin{itemize}
    \item Integer instructions with \gls{rs}\texttt{=x31} pop the operand from the F2I queue, instead of reading it from the integer \gls{rf};
    \item Integer instructions with \gls{rd}\texttt{=x31} push the operand to the I2F queue, instead of writing it to the integer \gls{rf};
    \item \gls{fp} instructions with integer \gls{rs} pop the operand from the I2F queue, instead of reading it from the integer \gls{rf};
    \item \gls{fp} instructions with integer \gls{rd} push the operand to the F2I queue, instead of writing it to the \gls{fp} \gls{rf}.
\end{itemize}

\begin{figure}[t]
    \begin{minipage}[t]{0.31\columnwidth}
        \vspace{0pt}
        \centering
        \begin{minipage}[t]{\textwidth}
            \begin{minted}{asm}
add t0, t0, t1
fcvt.d.wu ft0, t0
       ---
sw t1,  8(t0)
sw t2, 12(t0)
fld ft0, 8(t0)
fmul.d ft2, ft0, ft1
            \end{minted}
            \label{fig:base}
        \end{minipage}
    \end{minipage}%
    \hfill
    \textcolor{darkgray}{\vline}
    \hfill
    \begin{minipage}[t]{0.31\columnwidth}
        \vspace{0pt}
        \centering
        \begin{minipage}[t]{\textwidth}
            \begin{minted}[xleftmargin=-0pt]{asm}
add x31, t0, t1
fcvt.d.wu ft0, t0
       ---
sw t1,  8(t0)
sw t2, 12(t0)
addi x31, t0, 8
fld ft0, 0(t0)
fmul.d ft2, ft0, ft1
            \end{minted}
            \label{fig:base}
        \end{minipage}
    \end{minipage}%
    \hfill
    \textcolor{darkgray}{\vline}
    \hfill
    \begin{minipage}[t]{0.35\columnwidth}
        \vspace{0pt}
        \centering
        \begin{minipage}[t]{\textwidth}
            \begin{minted}{asm}
call configure_ft0_ssr
add t0, t0, t1
sw t0, 0(%[addr])
fcvt.d.wu.ssr ft3, ft0
          ---
call configure_ft0_ssr
sw t1,  8(t0)
sw t2, 12(t0)
fmul.d ft2, ft0, ft1
            \end{minted}
            \label{fig:chaining}
        \end{minipage}
    \end{minipage}
    \vspace{-0.4em}
    \caption{\Cref{step:queues} transform applied to register- (top) and memory-carried (bottom) dependencies: (left) original code; (center) COPIFTv2 transform; (right) equivalent COPIFT transform.}
    \label{fig:step-queues}
\end{figure}

Leveraging the queues, we can greatly simplify the COPIFT methodology: the complex software pipelining and multiple-buffering schemes (Steps 5 and 6) of the original COPIFT method are no longer required.
The updated COPIFTv2 methodology consists of the following steps:
\begin{enumerate}[label=\textbf{Step \arabic*},ref=\arabic*,leftmargin=4em]
    \item \label[step]{step:dfg} Construct a \gls{dfg} and identify all dependencies between integer and \gls{fp} instructions.
    \item \label[step]{step:partitioning} Partition the \gls{dfg} into a subgraph of integer-only and a subgraph of \gls{fp}-only instructions.
    \item \label[step]{step:scheduling} Schedule instructions in each subgraph to maximize the overlap between the two subgraphs.
    \item \label[step]{step:queues} Map all inter-thread communication to the queues.
    \item \label[step]{step:frep} Map the \gls{fp} subgraph to an \texttt{FREP} hardware loop.
\end{enumerate}
Unlike COPIFT, COPIFTv2 requires no loop transformations---all transformations are confined to the loop body---bringing the methodology closer to a form that a compiler can automate \cite{lopoukhine2025}.

For brevity, we focus only on \Cref{step:queues}, noting that \Cref{step:dfg,step:partitioning,step:scheduling,step:frep} are simple variations of corresponding steps in COPIFT.

\Cref{fig:step-queues} illustrates how to apply \Cref{step:queues} to inter-thread communication occurring through both the \gls{rf} and through memory.
In the first case, it is sufficient to substitute the register carrying the data dependency with \texttt{x31} in the integer thread, while the \gls{fp} instruction remains unchanged.
The same applies to communication in the other direction, from a \gls{fp} to an integer instruction.
Communication through memory, on the other hand, entails two dependencies: 1) the actual data dependency through memory, and 2) a register-carried dependency on the address (through \texttt{t0}, in this example).
As the offset calculation cannot be resolved in the \gls{fpss}, it must be performed in the integer thread and the resolved address passed through the I2F queue.
Mapping the address dependency to the queue also ensures that the ordering of the memory accesses is preserved.

\section{Results}

\begin{figure}[t]
    \begin{subfigure}{\columnwidth}
        \centering
        \includegraphics[width=\textwidth]{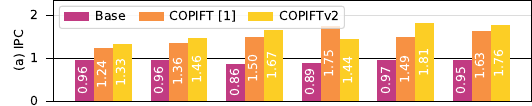}
        \phantomcaption
        \label{fig:ipc}
    \end{subfigure}
    \vspace{-1.1em}

    \begin{subfigure}{\columnwidth}
        \centering
        \includegraphics[width=\textwidth]{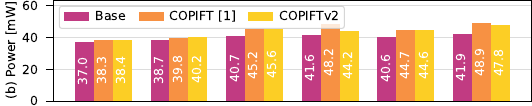}
        \phantomcaption
        \label{fig:power}
    \end{subfigure}
    \vspace{-1.2em}

    \begin{subfigure}{\columnwidth}
        \centering
        \includegraphics[width=\textwidth]{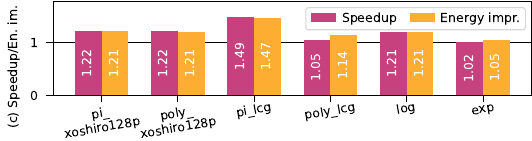}
        \phantomcaption
        \label{fig:speedup}
    \end{subfigure}
    \vspace{-1.4em}

    \caption{(a, b) Comparison of baseline, COPIFT and COPIFTv2. (c) Speedup and energy savings of COPIFTv2 over COPIFT.}
    \label{fig:results}
\end{figure}

We implement a Snitch cluster \cite{zaruba2021} with one compute core in GlobalFoundries' 12LP+ FinFET technology using Fusion Compiler 2023.12, with a target clock frequency of 1 GHz.
All experiments are conducted in cycle-accurate RTL simulations using QuestaSim 2023.4. Switching activities are extracted from post-layout simulations, and used for power estimation in PrimeTime 2022.03, assuming typical operating conditions of 25 °C and 0.8 V supply voltage.
Our extensions do not affect the critical path and introduce a negligible $<$1\% area overhead.


We evaluate COPIFTv2 on a set of mixed integer and \gls{fp} codes presented in \cite{colagrande2025}.
As shown in \Cref{fig:ipc}, COPIFTv2 achieves an \gls{ipc} improvement over all COPIFT codes, reaching a peak \gls{ipc} of \ResultPeakIPC, with exception of the \texttt{poly\_lcg} kernel, where COPIFT's overhead load/store instructions contribute to balance the integer and \gls{fp} threads.
These, however, do not contribute useful work, thus COPIFTv2 still achieves a higher overall throughput (samples/cycle) than COPIFT on all benchmarks, as shown in \Cref{fig:speedup}, resulting in \ResultPeakSpeedup$\times$ maximum and \ResultGeomeanSpeedup$\times$ geomean speedups over COPIFT.
Power consumption remains comparable between COPIFT and COPIFTv2, as shown in \Cref{fig:power}, with two opposing effects balancing each other: the increased \gls{ipc} tends to increase power, while communication through the queues instead of memory decreases it.
The similar power consumption, combined with the increased throughput, results in significant energy efficiency gains, as shown in \Cref{fig:speedup}, reaching up to \ResultPeakEnergySaving$\times$ maximum and \ResultGeomeanEnergySaving$\times$ geomean energy efficiency improvements over COPIFT.

\section{Conclusion}

In this work, we presented COPIFTv2, an enhanced dual-issue execution methodology which eliminates much of the complexity and software overheads of COPIFT, achieving up to a \ResultPeakSpeedup$\times$ speedup and a \ResultPeakEnergySaving$\times$ energy-efficiency gain over COPIFT, respectively \ResultPeakSpeedupOverSnitch$\times$ and \ResultPeakEnergySavingOverSnitch$\times$ over the Snitch baseline, and a peak \gls{ipc} of \ResultPeakIPC.
Overall, these results demonstrate that modest architectural support is sufficient to unlock targeted dual-issue performance without compromising energy-efficiency or programmability.

\bibliography{paper}
\bibliographystyle{IEEEtran}

\end{document}